\documentclass[9pt,twocolumn,twoside]{pnas-new}
\templatetype{pnasresearcharticle} % Choose template
\makeatletter
\def\PNAS@linecountLO{}
\def\PNAS@linecountRO{}
\def\PNAS@linecountLE{}
\def\PNAS@linecountRE{}

\def\PNAS@linecountsecpageLO{}
\def\PNAS@linecountsecpageRO{}
\def\PNAS@linecountsecpageLE{}
\def\PNAS@linecountsecpageRE{}
\makeatother

\fancypagestyle{firststyle}{\fancyhead{}\fancyfoot{}} % Clears headers/footers on firststyle
\usepackage{graphicx}% Include figure files
\usepackage{dcolumn}% Align table columns on decimal point

\usepackage{color}
\usepackage{bm}
\usepackage{ amssymb, amsmath }

\newcommand{\comm}[1]{#1} %commenti
\renewcommand{\AA}{\bm{A} }
\newcommand{\BB}{\bm{B} }

\newcommand{\VV}{\bm{V} }

\renewcommand{\aa}{\bm{a} }
\newcommand{\bb}{\bm{b} }

\newcommand{\kk}{\bm{k} }
\newcommand{\rr}{\bm{r} }

\newcommand{\xx}{\bm{x} }

\begin{document}

\title{Turbulence in the terrestrial magnetosheath: space-time correlation using the Magnetospheric Multiscale mission}

% Use letters for affiliations, numbers to show equal authorship (if applicable) and to indicate the corresponding author
\author[a,1]{Francesco Pecora}
% 0000-0003-4168-590X
\author[a]{William H. Matthaeus}
% whm@udel.edu
% 0000-0001-7224-6024
\author[b]{Antonella Greco}
% antonella.greco@fis.unical.it
% 0000-0001-5680-4487
\author[c,d]{Pablo Dmitruk}
% pdmitruk@df.uba.ar
% 0000-0002-7810-3451
\author[a]{Yan Yang}
% 0000-0003-2965-7906
% yanyang@udel.edu
\author[b]{Vincenzo Carbone}
% vincenzo.carbone@fis.unical.it
% 0000-0002-3182-6679
\author[b]{Sergio Servidio}
% sergio.servidio@fis.unical.it 
% 0000-0001-8184-2151

\affil[a]{Department of Physics and Astronomy, University of Delaware, Newark, DE 19716, USA}
\affil[b]{Dipartimento di Fisica, Universit\`a della Calabria, Arcavacata di Rende, 87036, IT}
\affil[c]{Universidad de Buenos Aires, Facultad de Ciencias Exactas y Naturales, Departamento de Física, Buenos Aires, Argentina}
\affil[d]{CONICET - Instituto de Física Interdisciplinaria y Aplicada (INFINA), Buenos Aires, Argentina}

% Please give the surname of the lead author for the running footer
\leadauthor{Pecora}

% Please add a significance statement to explain the relevance of your work
\significancestatement{This research examines chaotic behavior of magnetic field fluctuations within turbulent collisionless plasma, as observed by spacecraft in the geospace environment. A key challenge lies in disentangling spatial and temporal variations; spacecraft measurements struggle to differentiate whether observed changes are due to spatial decorrelation or due to intrinsic time evolution; this introduces fundamental uncertainty in data analysis. The study aims to determine what causes time-decorrelation between spatial points in the turbulent magnetosheath magnetic field: is it rapid sweeping of spatial structure, or strong local nonlinear interactions? Understanding these mechanisms is crucial for predicting how correlations persist in space-time. While similar problems are well-studied in hydrodynamics and magnetohydrodynamics, this fundamental problem remains largely unexplored in space plasma observation.
%Authors must submit a 120-word maximum statement about the significance of their research paper written at a level understandable to an undergraduate educated scientist outside their field of speciality. The primary goal of the significance statement is to explain the relevance of the work in broad context to a broad readership. The significance statement appears in the paper itself and is required for all research papers.
}

% Please include corresponding author, author contribution and author declaration information
\authorcontributions{Author contributions: F. Pecora, analysis, writing, algorithm development; W. Matthaeus, conceptualization, theory, writing; A. Greco, P Dmitruk, Y. Yang. V. Carbone, S. Servidio, preliminary and background research, contributions to conceptualization.}
\authordeclaration{Please declare any competing interests here. None.}
%\equalauthors{\textsuperscript{1}A.O.(Author One) contributed equally to this work with A.T. (Author Two) (remove if not applicable).}
\correspondingauthor{\textsuperscript{1}To whom correspondence should be addressed. E-mail: whm@udel.edu; fpecora@udel.edu}

% At least three keywords are required at submission. Please provide three to five keywords, separated by the pipe symbol.

\begin{abstract}
Spatiotemporal correlation of magnetic field fluctuations is investigated using the Magnetospheric Multiscale mission in the terrestrial magnetosheath. The first observation of the turbulence propagator in space emerges through analysis of more than a thousand intervals.
Results show clear features of spatial and spectral anisotropy, leading to a distinct behavior of relaxation times in the directions parallel and perpendicular to the mean magnetic field.
Full space-time investigation of the Taylor hypothesis reveals a scale-dependent anisotropy of magnetosheath fluctuations that can be compared to the effect of flow propagation on spacecraft frame time decorrelation rates as well as with Eulerian estimates.
The turbulence propagator reveals that the amplitudes of the perpendicular modes decorrelate according to sweeping or Alfvénic propagation mechanisms. The decorrelation time of parallel modes instead does not depend on the parallel wavenumber, which could be due to resonant interactions.
Through direct observation, this study provides unprecedented insight into the space-time structure of turbulent space plasmas, while giving critical constraints for theoretical and numerical models.
\end{abstract}

\dates{This manuscript was compiled on \today}
\doi{\url{www.pnas.org/cgi/doi/10.1073/pnas.XXXXXXXXXX}}

\maketitle
\thispagestyle{firststyle}
\ifthenelse{\boolean{shortarticle}}{\ifthenelse{\boolean{singlecolumn}}{\abscontentformatted}{\abscontent}}{}

\firstpage[1]{4} %#@ %\firstpage[line]{paragraph} %#@
% Use \firstpage to indicate which paragraph and line will start the second page and subsequent formatting. In this example, there are a total of 11 paragraphs on the first page, counting the first level heading as a paragraph. The value {12} represents the number of the paragraph starting the second page. If a paragraph runs over onto the second page, include a bracket with the paragraph line number starting the second page, followed by the paragraph number in curly brackets, e.g. "\firstpage[4]{11}".

% If your first paragraph (i.e. with the \dropcap) contains a list environment (quote, quotation, theorem, definition, enumerate, itemize...), the line after the list may have some extra indentation. If this is the case, add \parshape=0 to the end of the list environment.
\dropcap{T}he complex nature of turbulent flows is (partially) revealed when both spatial and temporal variations are considered. Spatial variations involve intricate structures such as vortices and eddies spanning a wide range of scales, from the smallest dissipation scales to the largest energy-containing scales. Instead, temporal variations reveal the evolution of dynamic processes such as intermittent bursts of activity. However, disentangling space and time is no easy task, especially when spacecraft measurements are involved.

A single spacecraft in the solar wind can measure single-point two-time correlations, generally dominated by the sweeping mechanism \citep{chen1989sweeping}. This refers to the dominance of advection of small-scale turbulent fluctuations by the larger-scale flow over intrinsic local dynamics in measurements. This is particularly relevant in high-speed flows like the solar wind, where large-scale structures carry smaller-scale variations past the spacecraft at a pace faster than their typical dynamical evolution time \citep{tennekes1975eulerian}. This is the basis of the Taylor frozen-in approximation \citep{taylor1938spectrum_frozenin} that assumes the temporal changes are dominated by the advection of spatial structures past the observer, neglecting intrinsic time variations. This allows for the conversion from frequency to wavenumber spectra, facilitating the analysis of turbulent structures in terms of spatial properties. Failures of this assumption are related, for example, to flows with low speeds or highly dynamic turbulence. Current measurements from Parker Solar Probe \citep{fox2016solar} identified scenarios where deviations from the Taylor hypothesis are likely to occur \citep{klein2015modified,bourouaine2018limitations,chhiber2019contextual,perez2021applicability}.

\comm{Another type of observation, one that is essentially impossible to achieve, is one in which the point of observation, i.e., the spacecraft, has a fixed position in the frame moving with the mean solar wind speed. This is the ``zero net momentum'' frame, analogous to a probe at a fixed position in a laboratory experiment. (The spacecraft motion is slow and generally negligible.) In this case, pure Eulerian decorrelations can be observed, and power spectra can be cast in frequency rather than wavenumber. This may be approximated with certain assumptions \citep{matthaeus2010eulerian}. In general, single-point measurements record a trajectory in space-time that is neither purely Eulerian nor purely Lagrangian, and the correlation function obtained by a single spacecraft measurement is composed of both spatial and temporal decorrelations.}

In favorable conditions, multiple spacecraft configurations can provide a (limited) coverage of the space-time domain as we present below using the 4-spacecraft Magnetospheric Multiscale (MMS) mission \citep{burch2016MMS}. In the future, this approach can be extended to the multiscale 9-spacecraft HelioSwarm mission in the solar wind \citep{klein2023helioswarm}, and to 7-spacecraft Plasma Observatory mission in the magnetosheath \citep{retino2022particle_PO,marcucci2024PO}. The relevance of this topic is found in the numerous theoretical and practical applications in turbulence theory \citep{batchelor1953theory,MoninYaglomI,MoninYaglomII,zhou2004colloquium}, particle scattering \citep{bieber1994proton,shalchi2006parallel}, and space weather predictions \citep{ridley2000estimations}.

The complete investigation of spatial and temporal correlations requires knowledge of the two-point two-time correlation function defined, for the magnetic field, as
\begin{equation}
    R(\rr,\tau) = \langle \bb(\xx+\rr, t+\tau) \cdot \bb(\xx,t) \rangle
    \label{eq:Rrt}
\end{equation}
\comm{with $\bb=\BB-\BB_0$ denoting magnetic field fluctuations about a mean field $\BB_0$, and $\langle . \rangle$ denoting an ensemble average.
The latter is generally interpreted as averaging over a suitably large domain, amounting to an assumption of an ergodic property \cite{orszag1977lectures}. For spacecraft observations, this is typically implemented by averaging over samples containing many correlation times (or lengths). In this case, this is achieved by selecting intervals with durations that satisfy this requirement \citep{roy2022turbulent}. The system is considered homogeneous and sampled over a volume large enough that Eq.~\ref{eq:Rrt} does not depend on the initial position $\xx$ nor the specific initial time $t$, but only on the space and time lags $\rr$ and $\tau$.}

Exploiting properties of multi-spacecraft missions, the spatial and temporal increments are investigated considering separate pairs of spacecraft, and Eq.~\ref{eq:Rrt} can be recast in terms of the $i$--$j$th pair of spacecraft, at positions $\xx^i$ and $\xx^j$, as
\begin{equation}
    R^{ij}(\rr,\tau) = \langle \bb(\xx^{i} - \VV\tau, t+\tau) \cdot \bb(\xx^{j},t) \rangle
    \label{eq:Rrtij}
\end{equation}
with $\bb^i$ and $\bb^j$ magnetic field measurements of the two spacecraft. The spatial increment in Eq.~\ref{eq:Rrtij} is $\rr = \xx^{i} - \VV\tau - \xx^j$, and the term $\VV\tau$ shifts the increments in the plasma rest frame with average velocity $\VV$ \comm{determined by averaging the flow speed over the duration of each interval. The spacecraft positions are assumed to be fixed since their relative speed is negligible (a few m/s).}
A fundamental property of the spatial correlation function is that it forms a Fourier pair with the power spectral density $S(\kk) = \int R(\rr) \, e^{-i\kk\cdot\rr} \, d\rr$  \citep{batchelor1953theory}. Analogously, the spatial Fourier transform of the two-point two-time correlation function is the time-lagged spectral energy density $S(\kk,\tau) = \int R(\rr,\tau) \, e^{-i\kk\cdot\rr} \, d\rr$ that can be factorized as
\begin{equation}
    S(\kk, \tau) = S(\kk) \Gamma(\kk, \tau).
    \label{eq:prop}
\end{equation}
\comm{This defines the propagator $\Gamma(\kk,\tau)$}, the scale-dependent function that describes the time decorrelation of spectral amplitudes. \comm{The factorization in Eq.~\ref{eq:prop} quantifies the decorrelation time as a function of scale $\Gamma (\kk,\tau)$ without regard to the spectral amplitude $S(\kk)$.} This quantity is central in turbulence theories \cite{edwards1964statistical,orszag1970analytical,zhou2004colloquium} and its determination has been a primary goal since the earliest days of numerical turbulence simulation \citep{orszag1972numerical}.

\comm{Here, we began by defining the propagator in terms of the magnetic field power spectral density as we delve into magnetohydrodynamics (MHD) and space plasma applications. However, the bulk of fundamental theoretical, numerical, and experimental developments were performed in hydrodynamics (HD). See \cite{zhou2021turbulence} for a review.}
Well studied in HD experiments \citep{comtebellot1971simple}, and MHD simulations \citep{servidio2011time,lugones2016spatiotemporal,lugones2019spatiotemporal}, until now, the propagator has yet to be estimated in space plasma, as far as we are aware.

The propagator $\Gamma$ in Eq. \ref{eq:prop}can be modeled as
\begin{equation}
    \Gamma = r[\gamma(\kk) \tau],
\label{eq:Gamma}
\end{equation}
\comm{using a generic function $r$ that models the decaying profile of the propagator in terms of scale-dependent decorrelation rates $\gamma(\kk)$. We are interested in the contribution of underlying physical processes to the total decorrelation rate $\gamma(\kk)$ in terms of (reciprocal) $\kk$-dependent timescales.}
\comm{Heuristically, the dependence of the decorrelation time on the wavenumber can be understood as follows. Larger structures (smaller wavenumbers $k$) can be imagined to have slower dynamical evolution time and to be subject to fewer modifications. In contrast, smaller structures (large $k$'s) undergo fast evolution and thus ``decorrelate'' faster. The observed scaling of the decorrelation times on the wavenumber can distinguish the physical processes driving the evolution of turbulence. Nonlinear distortions lead to time decorrelation rates depending on $\tau_{_{NL}} \sim k^{-2/3}$, while sweeping decorrelation by large-scale flows at speed $U$ depends on results $\tau_{sw} = 1/kU$, and sweeping by Alfv\'enic propagation depends on $\tau_A \sim 1/kV_A$ \cite{zhou2004colloquium}. Different phenomena produce other possible time scales, e.g., associated with dissipation, and the overall scaling can then be a combination of these \citep{comtebellot1971simple}.} 

If the decorrelation at scale $k$ is dominated by straining effects, the scaling of the decorrelation times is expected to be controlled by $\tau_{_{NL}}$. Kolmogorov's theory \citep{kolmogorov1941local} is based on the assumption of local-in-scale interactions. Energy enters the system at large scales and then cascades to smaller scales through interactions between neighboring scales via nonlinear straining and distortion of large vortices, fragmenting into smaller ones. The nonlinear time of the straining at scale $k$ is $\tau_{_{NL}}(k)~=~(ku_k)^{-1}$ where $u_k = \sqrt{k\hat{E}(k)}$ is the characteristic speed at wavenumber $k$. The omnidirectional (kinetic) energy spectrum $\hat{E}(k)$ is related to the spectral density as  $\hat{E}(k) = \int S(\kk) k^2 d\Omega$. Using the prescription $\tau_{_{NL}}(k)~\sim~k^{-2/3}$, Kolmogorov's inertial range scaling $\hat{E}(k) \sim k^{-5/3}$ is recovered.

The Alfvén propagation effect is analogous to sweeping in Iroshnikov \citep{iroshnikov1964turbulence} and Kraichnan \citep{kraichnan1965inertial} spectral transfer phenomenologies. In this case, nonlinear interactions are reduced in time due to the counter-propagation of Alfvén wave packets that are ``swept'' by the mean magnetic field, suppressing the spectral transfer rate by a factor $\tau_A/\tau_{_{NL}}(k)$ for Alfvén time $\tau_A~=~(kV_A)^{-1}$. \footnote{Sweeping in the context of the direct interaction approximation (DIA) theory \citep{kraichnan1959structure} gives rise to a $k^{-3/2}$ scaling of the power spectral density at large Reynolds number; however, this is corrected in the Galilean invariant version of the theory \citep{kraichnan1965lhdia}.} While spectral transfer rates are not our principal concern here, the role of Alfvénic sweeping in limiting the interactions is essentially a time decorrelation effect. Thus, when the interaction is dominated by velocity sweeping or Alfvénic propagation, the decorrelation time is supposed to scale with the inverse of the wavenumber.% In both cases, the $1/k$ scaling is expected.

Earlier numerical and experimental works provide supporting evidence for these scalings. Simulations of 3D isotropic HD turbulence at moderate Reynolds number \citep{orszag1972numerical, sanada1992random} show the propagator scaling with the inverse of the wavenumber as in sweeping-dominated turbulence. A varying Reynolds number study of 3D isotropic HD turbulence \citep{mccomb1989velocity} showed that neither pure sweeping nor pure straining provides an accurate scaling for the propagator. More accurate scaling is attained with the sweeping prediction at small Reynolds numbers, transitioning to straining scaling at large Reynolds numbers.

More recently, using 3D isotropic MHD simulations \citep{servidio2011time}, the presence of sweeping decorrelation was found to be dominant. However, there is evidence that different wavenumber ranges can be affected by different decorrelation mechanisms, also depending on the strength of a background magnetic field \citep{lugones2016spatiotemporal}.

\comm{Complex turbulent flows in HD, MHD, and plasma systems are more effectively described by a combination of decorrelation effects. The coexistence of multiple decorrelation rates has been observed in HD wind tunnel experiments \cite{comtebellot1971simple}, and in the analyses of turbulent flows generated by Rayleigh-Taylor and Richtmyer-Meshkov instabilities \citep{zhou2024hydrodynamic}. The challenge of identifying the dominant timescales and discerning the relevant decorrelation mechanisms has been at the core of a decades-long debate regarding the fundamental physics governing spectra and spectral transfer, particularly in space plasmas \citep{brunocarbone2005review}.}

%%%%%%%%%%%%%%%%%%%%%%%%%%%%%%%%%%%%%%%%%%%%%%%%%%%%%%%%%%%%%%%%%
%%%%%%%%%%%%%%%%%%%%%%%%%%%%%%%%%%%%%%%%%%%%%%%%%%%%%%%%%%%%%%%%%
%%%%%%%%%%%%%%%%%%%%%%%%%%%%%%%%%%%%%%%%%%%%%%%%%%%%%%%%%%%%%%%%%
To investigate space-time correlations for the first time in the magnetosheath, we delve into the turbulence correlation properties in real (lag) space and spectral domains.

\section*{Space-time correlation function}

We computed the two-point two-time correlation function, defined in Eq.~\ref{eq:Rrtij}, using MMS four-spacecraft data, generating six independent pair measurements and four zero initial lag correlation functions within each of the 1180 analyzed intervals. The magnetic field is rotated into a frame where the parallel direction is aligned with the mean magnetic field of each interval. \comm{The data sets and spacecraft orientations are described in the Data selection section in Materials and Methods.} Each correlation function populates a stripe along the characteristic $\VV\tau$ with $\VV$ the average flow speed in that interval.

\begin{figure}[ht]
    \centering
    \includegraphics[width=\linewidth]{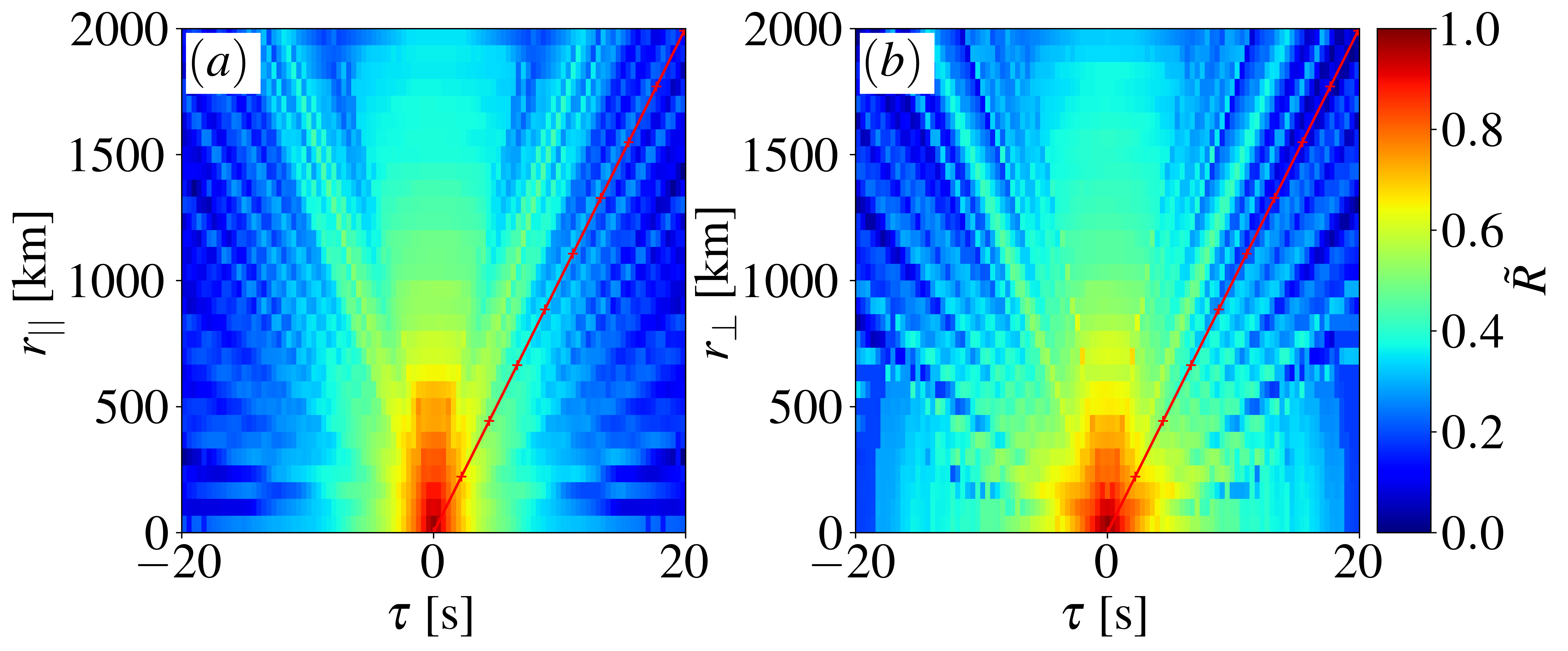}
    \caption{Space-time correlation function with complete coverage of the space-time domain. The red oblique line is the direction along a plasma parcel flowing at a speed of $100$~km/s.}
    \label{fig:R_fit_interp}
\end{figure}

The scarcity of very fast and slow plasma flows and extreme magnetic field angles prevents a complete sampling of the entire 4D space (3D spatial + time). For further analysis, we first reduce the correlation functions to a 3D (2D + time) domain, assuming axisymmetry relative to the mean magnetic field. That is, the 3D spatial vector increment is projected onto the parallel-perpendicular plane ($\mathbf{r} \mapsto r_\parallel, r_\perp$). Furthermore, we investigate separately the parallel and perpendicular correlations by integrating over the ignored direction $R(r_\perp, \tau) = \int dr_\parallel R(r_\parallel, r_\perp, \tau)$, and analogously for $R(r_\parallel, \tau)$. The effect of this on spectra will be discussed below. The Appendix illustrates our binning strategy and the number of data points per bin. It also details the method used to reconstruct the missing parts of the 2D (1D spatial + time) correlation functions. The resulting interpolated two-point two-time normalized correlation functions, $\tilde{R}(r,\tau) = R(r,\tau)/R(0,0)$, are presented in Fig.~\ref{fig:R_fit_interp}.

\comm{The spatiotemporal correlation functions in Fig.~\ref{fig:R_fit_interp} exhibit characteristic features \citep{batchelor1953theory,frisch1995turbulence}: The maximum correlation is observed at the origin of the lag space ($r = 0$, $\tau = 0$). As spatial and temporal separations grow, the correlations decay, approaching small values indicating decorrelation. Notably, the correlation is more persistent along the parallel spatial direction, resulting in elongated contours in the spatiotemporal correlation function. Conversely, the less elongated contours in the perpendicular plane suggest a more isotropic correlation in this plane.}

\section*{Real-space analysis -- Taylor frozen-in flow approximation}
Single-spacecraft measurements only allow for the investigation of single-point two-time correlations, which in the solar wind are primarily influenced by systematic sweeping effects as small perturbations are rapidly advected past the observer, potentially without significant evolution \citep{lumley1965interpretation,jokipii1973turbulence}. This is the core assumption of Taylor's frozen-in hypothesis \citep{taylor1938spectrum_frozenin}: the plasma flow speed past the spacecraft exceeds the dynamical evolution timescale, implying that observed temporal variations mainly reflect spatial structures. This commonly used principle allows for mapping spacecraft time series into spatial information.

\begin{figure}[ht]
    \centering
    \includegraphics[width=0.95\linewidth]{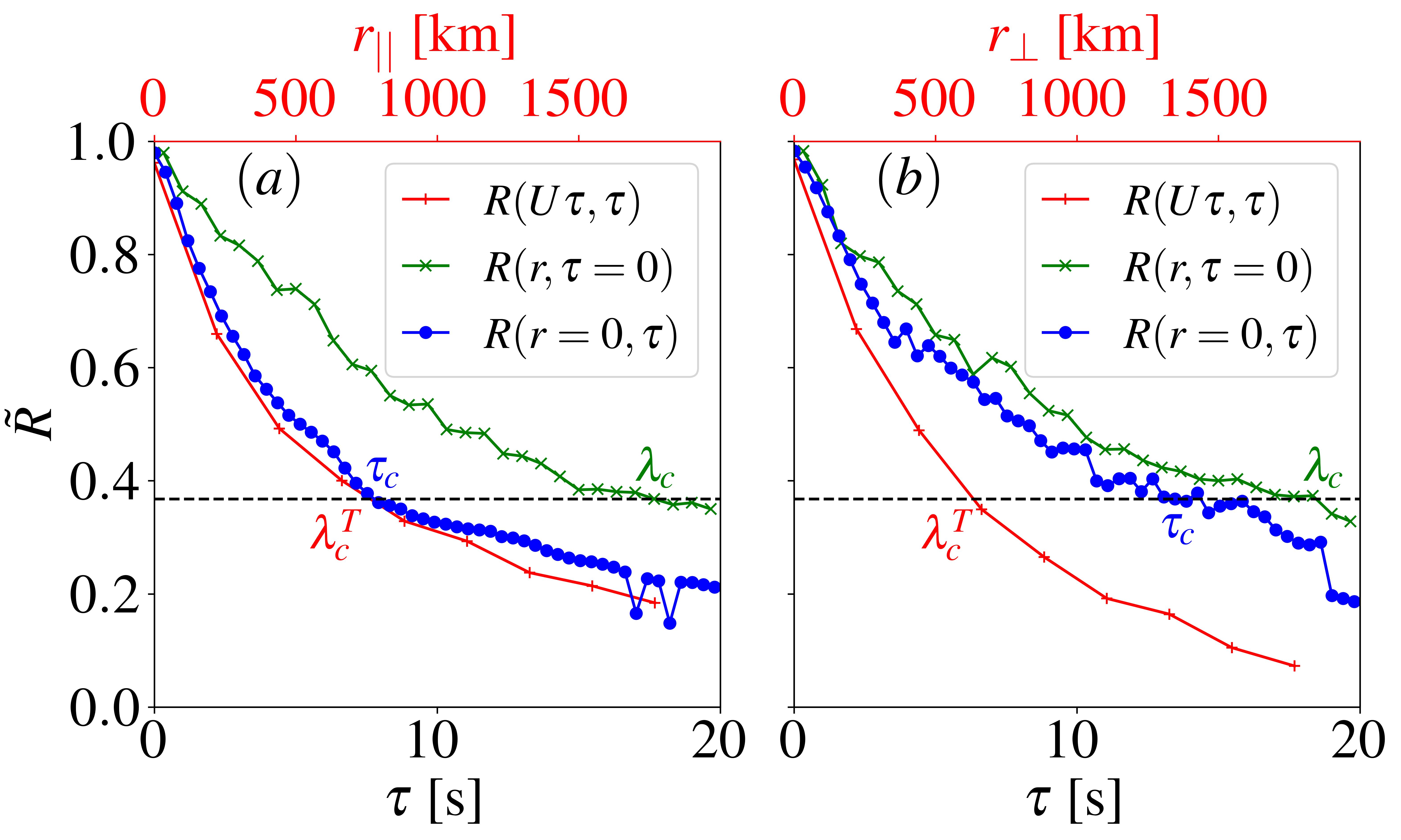}
    \caption{1D cuts of the normalized two-point two-time correlation function.
    The sample along a nominal 100 km/s flow speed is the red line with ``+'' symbols.
    The purely spatial (vertical) direction $\tau=0$ is the green line with ``x'' symbols. 
    The purely temporal (Eulerian, horizontal) sampling at $r=0$ is the blue line with circles.
    The horizontal line at $1/e$ marks where correlation times and lengths are estimated (annotated in the figure and reported in Table~\ref{tab:table1}).}
    \label{fig:R_1d}
\end{figure}

By examining the measured correlation function along three distinct directions in the space-time plane, we can challenge the validity of this assumption in the magnetosheath, similar to studies in the solar wind \cite{matthaeus2016ensemble}. Figure~\ref{fig:R_1d} displays three 1D cuts of the normalized correlation function. Referring to Figure~\ref{fig:R_fit_interp}, these slices correspond to horizontal, vertical, and oblique directions. The latter represents a plasma flow of 100~km/s – a typical speed in the magnetosheath \citep{pecora2025ubmsh}.

These 1D cuts allow the estimate of the correlation times/lengths in three different cases: (i) purely temporal $\tau_c$ (blue line), (ii) purely spatial domain $\lambda_c$ (green), and (iii) mixed space-time $\lambda_c^T$ (red). Table~\ref{tab:table1} reports the numerical values of these correlation scales obtained as the 1/e-folding of the normalized correlation function $\tilde{R}$.

\begin{table}[ht]
    \centering

    \setlength{\tabcolsep}{10pt} % Default value: 6pt -- column separation
    \renewcommand{\arraystretch}{1.3} % Default value: 1 -- row separation
    \begin{tabular}{ccc}
    \hline
    \hline
                           & Parallel & Perpendicular \\
                           \hline
        % 200 km/s
        % $\lambda_c^T$ [km] &  1268    &  1203         \\
        % $\lambda_c$ [km]   &  1845    &  1605         \\
        % $\tau_c$ [s]       &  8       &  11           \\
        %
        %100 km/s
        $\lambda_c^T$ [km] &  764    &  634         \\
        $\lambda_c$ [km]   &  1766   &  1845        \\
        $\tau_c$ [s]       &  7.8    &  13.4        \\
        
    \hline
    \hline
    \end{tabular}
    \caption{Correlation lengths and times obtained from the 1/e-folding of the 1D cuts in Fig.~\ref{fig:R_1d}. $\lambda_c^T$ is the correlation length along the characteristic with a nominal speed of 100~km/s (red line in Fig.~\ref{fig:R_1d}). $\lambda_c$ and $\tau_c$ are the correlation length and time in the purely spatial and temporal directions, respectively (green and blue lines in Fig.~\ref{fig:R_1d}).}
    \label{tab:table1}
\end{table}

The measured correlation lengths in the parallel direction are comparable to those in the perpendicular plane, suggesting isotropy at the outer scale. However, the correlations measured using Taylor's assumption favor stronger gradients in the perpendicular direction. This is suggestive of a flux tube-like pattern in which these structures are elongated in the parallel direction. In the pure time correlation, we observe the possibly unexpected result that correlation time is larger in the perpendicular direction. This may be symptomatic of inaccuracies in using Taylor's hypothesis. On the other hand, the sense of anisotropy measured from multidimensional correlation functions may vary not only in direction but also in scale, as seen in a recent study of density fluctuations at 1~au \cite{wang2024anisotropy}. Such behavior indicates that the highly anisotropic, field-aligned flux-tube scenario is likely not applicable at all scales. Indeed, the flux tube picture may require special conditions, such as low plasma $\beta$, a condition not often observed in the magnetosheath \cite{pecora2025ubmsh}.

\section*{Fourier-space analysis -- Scale dependent propagator}
To investigate the implications of the observed anisotropy properties, we now examine the turbulence correlator in the spectral domain. The normalized time-lagged spectral density $S(\kk, \tau)/S(\kk)$, Eq.~\ref{eq:prop}, is obtained by transforming the space-time autocorrelation function, Eq.~\ref{eq:Rrtij}, with $\kk$ the full 3D $k$-space vector. The procedure described above -- projection onto the parallel-perpendicular plane and integration along one coordinate -- reduces the power spectral density to an axisymmetric form, with either $k_\parallel=0$ or $k_\perp=0$. In the first case, $\int dr_\parallel e^{-ik_\parallel r_\parallel} \int dr_\perp e^{-ik_\perp r_\perp} R(r_\perp,\tau)  = S(0, k_\perp, \tau)$. This follows since the first integral is $2\pi\delta(k_\parallel)$ reducing $S(k_\parallel, k_\perp, \tau)$ to $S(0, k_\perp, \tau)$. Similarly, integrating over the perpendicular direction yields $S(k_\parallel, 0, \tau)$.

\begin{figure}[ht]
    \centering
    \includegraphics[width=\linewidth]{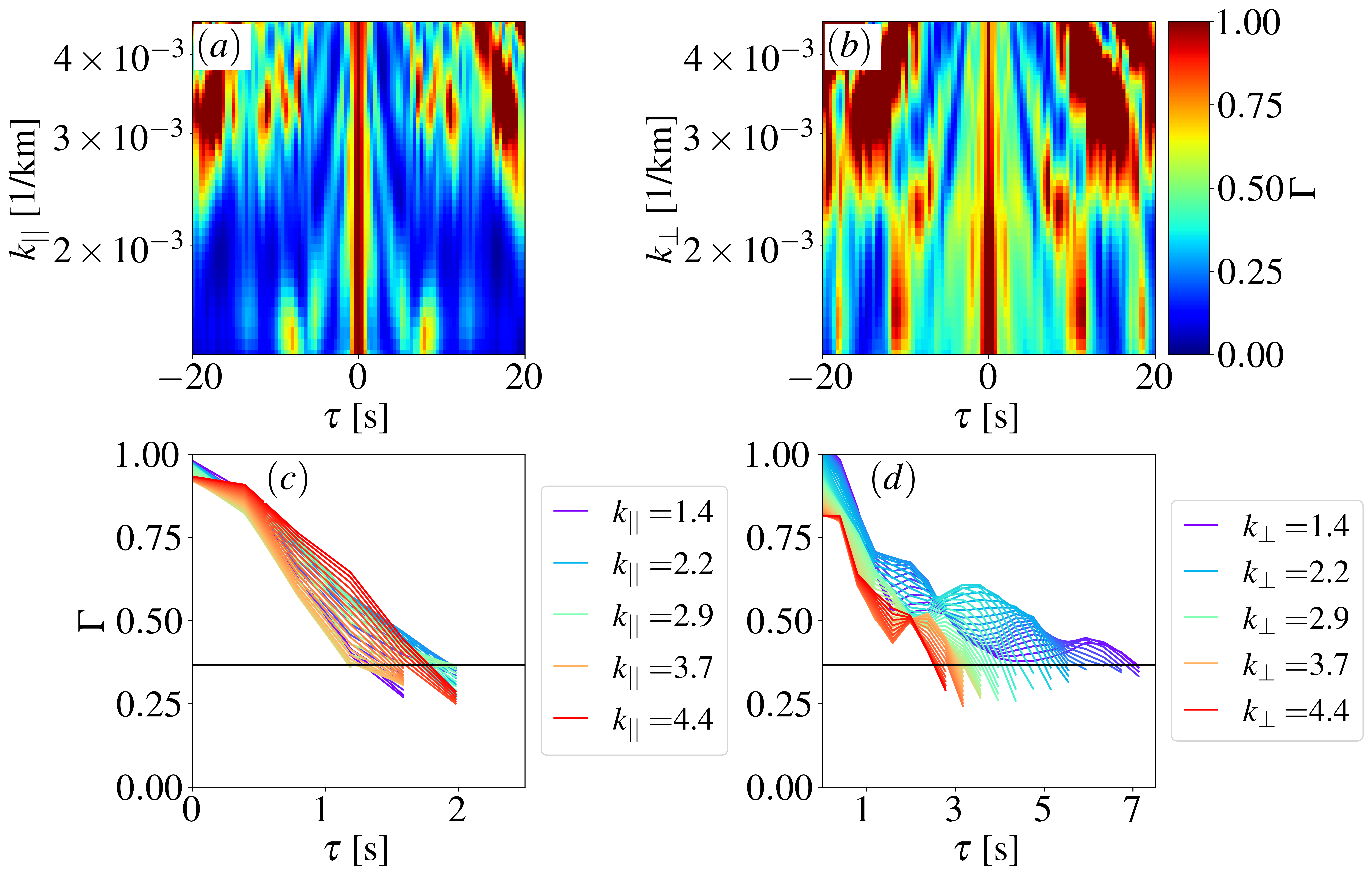}
    \caption{Decorrelator in the (a) $(k_\parallel,\tau)$ and (b) $(k_\perp,\tau)$ planes. (c), (d) Horizontal cuts of the decorrelator in the wavenumber range $1.3 \times 10^{-3}~\mathrm{km^{-1}} < k < 4.5 \times 10^{-3}~\mathrm{km^{-1}}$ for $k_\parallel$ and $k_\perp$ respectively. For each wavenumber, the correlation time is determined by the intercept with the $1/e$ horizontal line. For clarity, the lines are trimmed below the $1/e$ threshold. The labels report a few $k$ values to guide the reader through colors. $k$ is in units of $10^{-3}$ km$^{-1}$.}
    \label{fig:Skt}
\end{figure}

The top panels of Fig.~\ref{fig:Skt} show the decorrelators $\Gamma(k_\parallel,\tau)=S(k_\parallel,\tau)/S(k_\parallel)$ and $\Gamma(k_\perp,\tau)=S(k_\perp,\tau)/S(k_\perp)$. 

\comm{Although the decorrelator lacks a universally prescribed functional form, our observations indicate it is symmetric with respect to temporal lags. This property allows us to restrict the analysis to positive temporal lags ($\tau > 0$). While the detailed interpretation of the features within the full $(k, \tau)$ plane is deferred to future numerical investigations, we can gain initial insight into the decorrelator's behavior by reducing the dimensionality of the problem. This is achieved by taking horizontal cuts at fixed wavenumber $k$ and focusing on positive temporal lags.}

We determine the scale-dependent decorrelation by examining the profiles of $\Gamma(k,\tau)$ at fixed $k^*$'s (either $k_\parallel$ or $k_\perp$) as a function of $\tau$. The 1D cuts of $\Gamma(k,\tau)$ are shown in the bottom panels of Fig.~\ref{fig:Skt} for increasing positive $\tau$ (the behavior for negative increments is nearly identical). The decorrelation time for each scale is defined as the $1/e$ intercept of each $\Gamma(k^*,\tau)$.

\comm{For the presented analyses, we considered two types of error. First, to account for instrumental uncertainties, we introduced white noise to the magnetic field and plasma data, as done in \citep{pecora2023threedimensional_PRL}, where it was found to have no significant effect on the results. Second, we considered counting statistics. Since we limited our analysis to space-time regions with a minimum of $5 \times 10^5$ counts per bin, the statistical uncertainties are negligible. In both cases, the resulting error bars are smaller than the symbols shown in the figures.}

Here we focused on $1.3 \times 10^{-3}~\mathrm{km^{-1}} < k < 4.5 \times 10^{-3}~\mathrm{km^{-1}}$ or equivalently $ 200~\mathrm{km} < L < 750~\mathrm{km} $ corresponding to a few to several ion inertial lengths $d_i \sim 50$ km. These scales nominally belong in the turbulence inertial range, albeit its determination in the magnetosheath is not always accurate or possible due to the limited size of the system  \citep{huang2017existence,sahraoui2020magnetohydrodynamic,pecora2025ubmsh}.

\begin{figure}[ht]
    \centering
    \includegraphics[width=\linewidth]{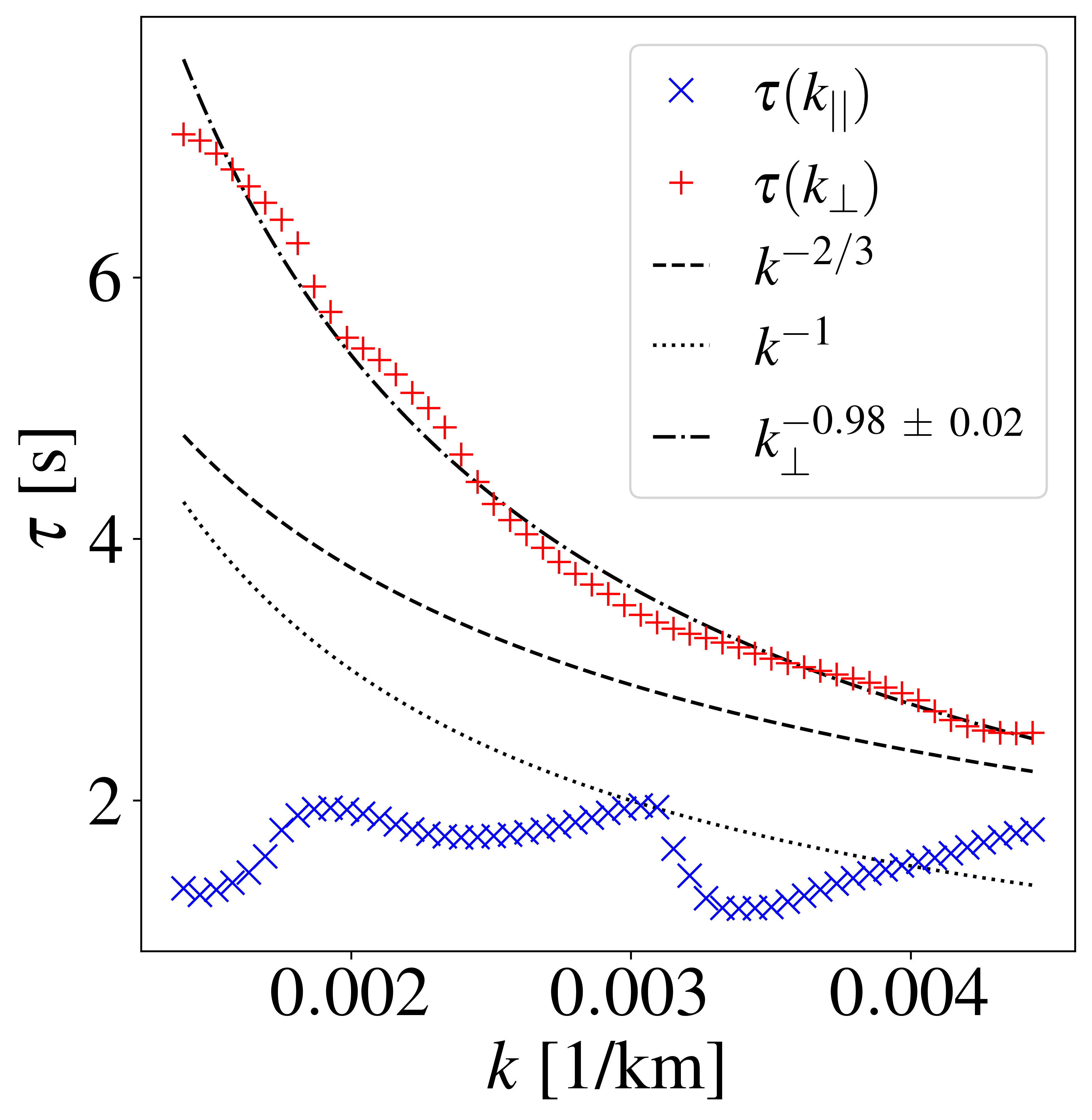}
    \caption{Scale-dependent time decorrelation of parallel and perpendicular wavenumbers. Blue ``x'' and red ``+'' symbols are the $1/e$-intercepts of $\Gamma$ profiles in Fig.~\ref{fig:Skt} for $k_\parallel$ and $k_\perp$ respectively. Reported are reference scalings for the nonlinear strain ($k^{-2/3}$ dashed line) and sweeping ($k^{-1}$ dotted line) predictions. 
    Parallel decorrelation times do not show dependence on the parallel wavenumber, suggesting independent decorrelation mechanisms. Perpendicular decorrelation times are fitted with a power law (dash-dotted line) providing a scaling exponent of $-0.98 \pm 0.02$ in excellent agreement with sweeping or Alfvénic theoretical predictions.}
    \label{fig:tauk}
\end{figure}

The determined scale-dependent decorrelation times over the selected range of scales are reported in Fig.~\ref{fig:tauk}. We note that the employed method does not distinguish between familiar functional forms for sweeping \ref{eq:Gamma}, such as exponential $r=e^{-\tau/\tau_0}$ or Gaussian $r=e^{-(\tau/\tau_0)^2}$ for arbitrary decorrelation time $\tau_0$. However, the method does clearly distinguish between cases where the controlling rate is $\tau_0 = \tau_{_{NL}}$ or $\tau_0 = \tau_{sw}$.

We observe distinct behaviors between parallel and perpendicular modes. The parallel modes show no notable dependence on the parallel wave vector, suggesting that decorrelation mechanisms along this direction operate independently from processes in the parallel direction. In contrast, the perpendicular direction reveals a much clearer pattern. Here, wave vectors decorrelate proportionally to $k^{-1}$, aligning with expectations from either sweeping decorrelation or Alfvénic propagation. This scaling relationship has been previously observed in analyses of isotropic MHD simulations \citep{servidio2011time,lugones2016spatiotemporal,lugones2019spatiotemporal}, but never before observed in plasma as far as we are aware.

\section*{Discussion}
The availability of multispacecraft observations by MMS allows for the first in-depth investigation of the space-time correlation in space plasma.

From the structure of the two-point two-time magnetic field correlation function, we can investigate the validity of Taylor's hypothesis. This often-invoked approximation relies on the assumption that the plasma flow is frozen during the observation time. Thus, in this approximation, all variations are purely spatial (corresponding to $\tau=0$ in the space-time domain). In the present analysis, the purely spatial decorrelation is separately evaluated so that the Taylor assumption can be examined for accuracy. The finding is that the frozen-in assumption is reasonably accurate for parallel lags, but is less so in the perpendicular direction.
%The orthogonal direction in the space-time plane at zero spatial lag provides the Eulerian two-time correlation function.

%When access to the full space-time ensemble is available, as in the present case, these measurements can be compared with the correlation time along a specific flow direction.
When the system is anisotropic, as observed here, there can be a discrepancy between the estimates of the correlation time and lengths deduced by these approaches. Of relevance is that in Taylor's approximation, the system seems more isotropic at the outer scale than observed along a characteristic flow speed of 100~km/s, with a tendency to favor flux-tube structures directed along the background magnetic field. The Eulerian and oblique observations, however, show a reversed trend with the perpendicular dimension being larger. This can be potentially attributed to a scale-dependent anisotropy as recently observed for density structures in the solar wind \citep{wang2024anisotropy}.

The space-time correlation function enables the study of scale-dependent decorrelation rates, here shown in a wavenumber decomposition. This quantifies a fundamental physical process that can be used to indirectly discriminate between alternative turbulence theories that purport to describe the cascade process. In fact, the two major theories, Kolmogorov's strain and Kraichnan's sweeping, predict different wavenumber power laws for the scale-dependent decorrelation times. The former describes the cascade process dominated by the straining effects between similar-in-size eddies which interact and form smaller scales producing the cascade. In this case, modes decorrelate as $k^{-2/3}$. In Kraichnan's development, instead, the dominant effect is the sweeping of smaller eddies by larger ones, and modes decorrelate as $k^{-1}$. In MHD, an identical scaling is obtained considering the nonlinear interaction of counterpropagating Alfvénic modes along the background magnetic field. In this case, modes also decorrelate as $k^{-1}$, as seen in related analytical developments \citep{azelis2024spacetime}.

%This theoretical expectation is the same as that expected when large eddies ``sweep'' the smaller eddies without distortion \citep{chen1989sweeping,zhou2004colloquium}.

In the terrestrial magnetosheath, we find that the perpendicular modes follow the sweeping or Alfvénic description. However, due to limitations in space-time coverage, it is not possible in our observations to distinguish between Alfvénic and sweeping decorrelation. Numerical investigations  \citep{servidio2011time,lugones2016spatiotemporal,lugones2019spatiotemporal} support the interpretation that the $1/k$ scaling observed in the perpendicular wavenumber space is attributable to sweeping decorrelation, and not to Alfv\'enic propagation. Lugones et al. \cite{lugones2016spatiotemporal} found that the $1/k$ scaling was mainly consistent with sweeping for all $k_\parallel$. There is a special case for which one may rule out Alfvénic propagation, namely the nonpropagating modes with $k_\parallel=0$ (as in this case when analyzing perpendicular wave vectors). In that case, all decorrelation is due to other effects, such as random sweeping or nonlinear distortion. As far as we are aware, of these, only the random sweeping accounts for $1/k$ scaling of the decorrelation times.

In the case of the Alfvénic mechanism, the decorrelation time explicitly depends on $k_\parallel$ only. However, in the present observations, decorrelation times of parallel modes do not seem to depend on $k_\parallel$. There is a well-defined theoretical development that may provide an explanation, although further studies are needed to draw a firm conclusion. It has been shown in both two- and three-dimensions \citep{shebalin1983anisotropy,oughton1994influence,oughton2006twocomponent} that {\it resonant interactions} dominate the dynamics of finite $k_\parallel$ modes in incompressible MHD turbulence in the presence of a guide field. This implies that the leading order triadic nonlinear couplings that drive these modes towards small scales involve only (near-) zero frequency turbulence in quasi-two-dimensional modes. The result is spectral transfer in the perpendicular wavenumber direction. 

Such a cascade conserves wave frequency, and the nonlinear cascade rate of the affected nonzero $k_\parallel$ modes depends only on the $k_\perp$ structure of the quasi-2D modes. In this process, again at leading order, one expects no dependence of decorrelation times on $k_\parallel$. In this perspective, the present observation may indicate that the parallel ``slab-like'' modes in the magnetosheath are decorrelating in time due to resonant couplings with zero-frequency eddies and nonpropagating structures. As far as we know, this is the first observation in space plasma that tests and supports such a hypothesis.  
\comm{Further research will be needed to understand the general conditions that cause suppression of nonresonant couplings, and in particular to understand this effect in the magnetosheath.}

\comm{There are outstanding questions regarding the applicability of the present results to space plasma other than the magnetosheath, and interesting prospects for addressing them. In the solar wind, a larger system size and a different range of parameters, such as mean field strength and plasma beta, become accessible. Furthermore, in the solar wind, the high-speed, super-Alfvénic nature of its flows motivates the expectation that decorrelation is primarily caused by the advection of turbulent structures. Sweeping effect is the dominant mechanism when the bulk flow speed is much greater than the characteristic times of, e.g., nonlinear interactions, and the Taylor hypothesis then becomes more accurate. Future multipoint and (especially) multiscale missions such as Plasma Observatory \citep{retino2022particle_PO,marcucci2024PO} and HelioSwarm \citep{klein2023helioswarm}, with appropriately spaced spacecraft, will provide a new perspective. With multiple vantage points and, in particular, by observing the solar wind when its flow is aligned with the separation vector of spacecraft, one might potentially isolate and observe nonlinear contributions to decorrelation effects distinct from the sweeping effect.}

\matmethods{

\subsection*{Data selection}
\comm{Figure~\ref{fig:mpause_intervals} illustrates the spatial distribution of the considered 1180 magnetosheath intervals, denoted by green symbols, in relation to a nominal magnetopause boundary. This boundary is defined using the model developed by \cite{shue1997new}. The Earth is represented by a blue circle. The inset provides a three-dimensional view of a typical tetrahedral configuration of the MMS constellation (here, corresponding to the interval whose symbol is highlighted in red right at the nose of the magnetopause on 2018 Feb 17, 22:08:43).
Geocentric Solar Ecliptic (GSE) coordinates are used, where x is directed from Earth to Sun, y is in the ecliptic plane pointing towards dusk (opposing planetary motion), and z is parallel to the ecliptic pole.}
We employ magnetic field data from the fluxgate magnetometers \cite[FGM,][]{russell2016magnetospheric} aboard the four-spacecraft Magnetospheric Multiscale mission (MMS) \citep{burch2016MMS}. Magnetic field measurements, available at 128~Hz in burst mode, are resampled to a grid uniformly spaced by 0.15~s as we are not interested in the high-frequency signatures. The data set consists of 1180 intervals of the turbulent magnetosheath between 2015 September 8 and 2018 January 1. These intervals are long enough so that a single-spacecraft correlation time can be properly evaluated in each of them \citep{roy2022turbulent}. \comm{The duration of these intervals ranges from 50 to 540 s, providing a cumulative dataset of $\sim 1.6\times 10^5$ s ($\sim 46$ hours). A distribution of the durations of these intervals is available in \citep{pecora2023relaxation}. These intervals are characterized by average ion inertial length $d_i=50$ km, corresponding to a plasma frequency $\omega_{pi}=6 10^3$ Hz. The average correlation time is $\tau_c = 10$ s or 2000 km using Taylor's hypothesis with an average flow speed of 200 km/s in the negative x GSE direction.}

\begin{figure}[ht]
    \centering
    \includegraphics[width=0.8\linewidth]{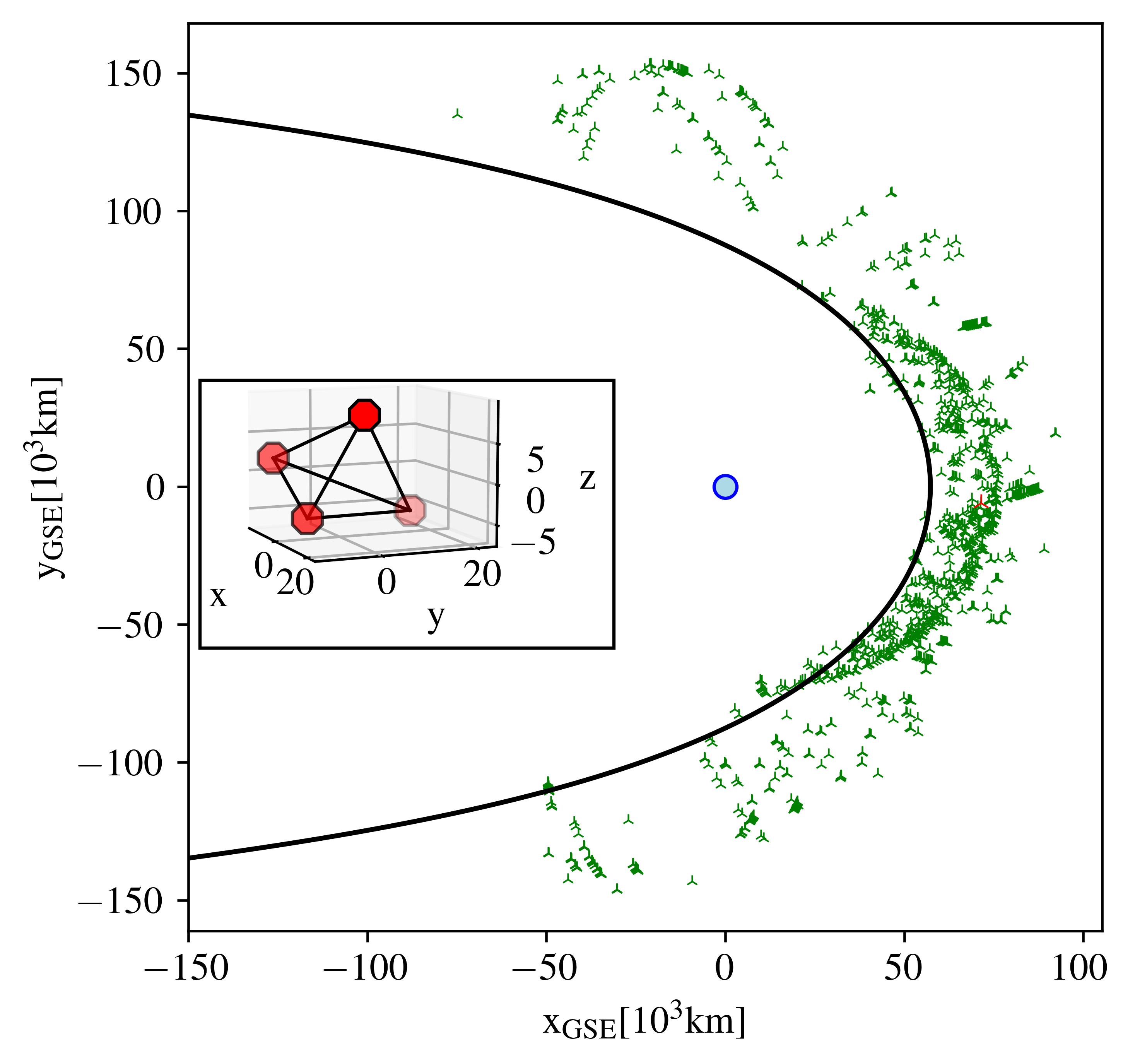}
    \caption{Position of the 1180 MMS magnetosheath intervals (green symbols). Nominal magnetopause boundary indicated with a thick black line. Earth is depicted as a blue circle. The inset shows the MMS constellation on 2018 Feb 17, 22:08:43, corresponding to the interval highlighted with a red symbol. The units of position in the inset are in km relative to the position of the barycenter.}
    \label{fig:mpause_intervals}
\end{figure}

\subsection*{Space-time coverage}
Figure~\ref{fig:Rrt} illustrates the binning domain in the reduced space-time plane, with panels (a) and (b) showing the normalized correlation function $\tilde{R} = R / R(0,0)$ in the parallel and perpendicular directions, respectively. We also count the number of points contributing to each bin, visualized in panels (c) and (d). White regions in (c) and (d) are areas of missing data coverage due to the absence of extremely fast or slow flows. We also noticed that bins with fewer than $5 \times 10^5$ counts manifested unphysical behavior, such as correlation function values anomalously increasing farther from the origin. Thus, we excluded these bins. This is reflected in the larger proportion of white areas in (a) and (b) compared to (c) and (d). This approach ensures that our analyses rely only on statistically robust and physically meaningful data.

\begin{figure}[ht]
    \centering
    \includegraphics[width=\linewidth]{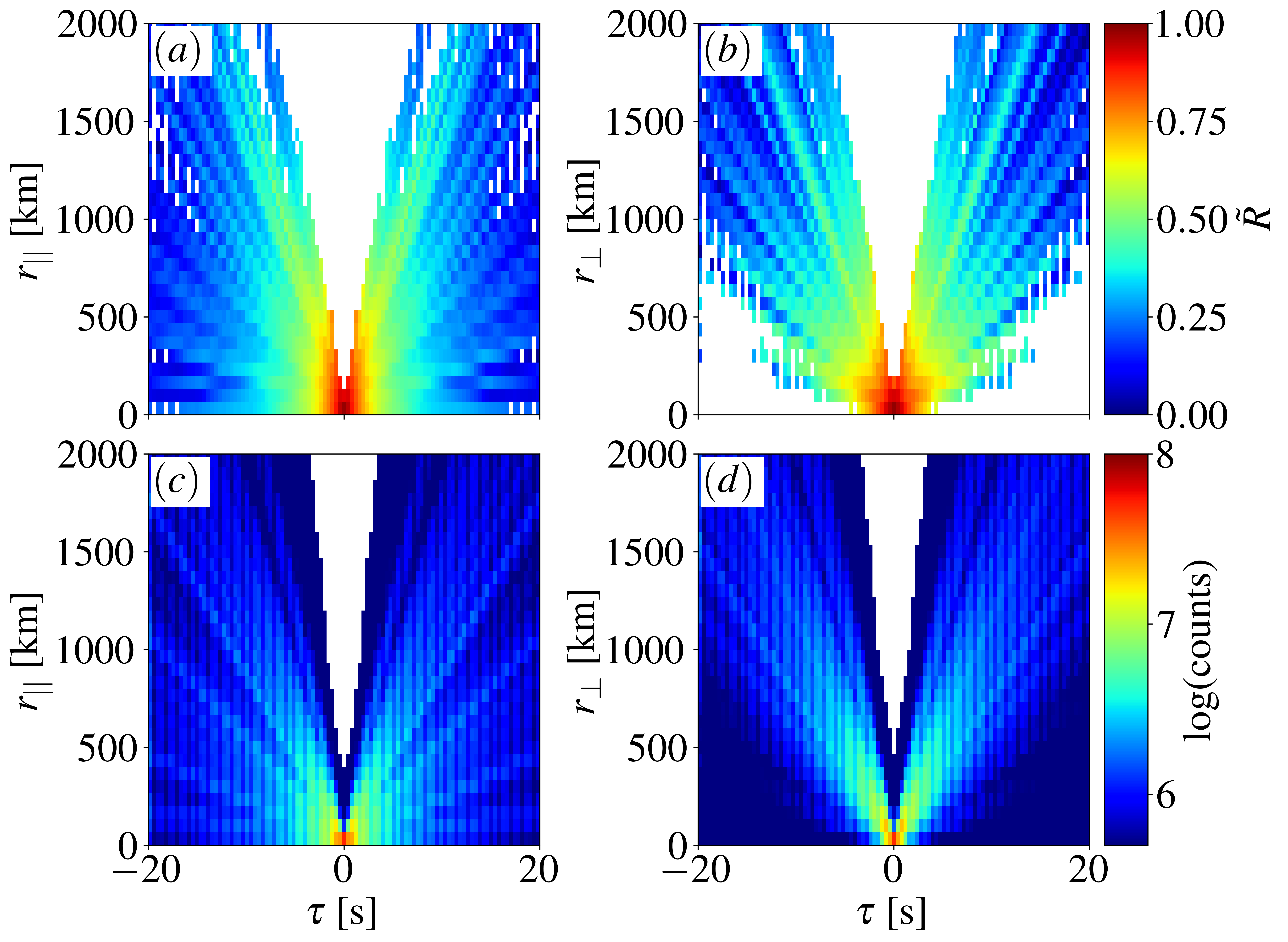}
    \caption{(a), (b) Space-time correlation function in the directions parallel and perpendicular to the mean field. White areas are missing coverage or low statistical counts. (c), (d) Counts in each bin of the space-time domain. Bins with counts less than $5\times 10^5$ are excluded from analyses.}
    \label{fig:Rrt}
\end{figure}

\subsection*{Filling the gaps}

To perform our analyses, we process the data to attempt complete coverage of the space-time domain. The binned correlation functions, obtained on a rectangular grid as in Fig.~\ref{fig:Rrt}~(a), (b), are uniformly sampled along curved paths, indicated by the black lines in Fig.~\ref{fig:fit}~(a) and (d), centered at the origin of the plane and parametrized via an angle $\theta$. Each point on the sampling curve has coordinates $[\tau \cos(\theta), r \sin(\theta)]$ (when not specified, $r$ is either $r_\parallel$ or $r_\perp$) with $\theta \in [ 0,\pi ]$. On each curve, 120 points are collected. To increase the number of points, we exploit the evenness property of the correlation functions ($R(\rr,\tau)=R(-\rr,-\tau)$) \citep{batchelor1953theory}. In this way, extending $\theta \in [ 0, 2 \pi )$, we obtain a periodic function of $\theta$ with sparse gaps.

The goal is to obtain complete coverage of the phase space by fitting and interpolating the measured values of the correlation function, assuming that no discontinuities are present. The sampled values of the correlation function along the parametric curves are periodic and thus can be properly fitted with the following series
\begin{equation}
    f(\theta) = \sum_{m=1}^5 f_m + A_m \cos(\theta_m + \phi_m).
    \label{eq:fit}
\end{equation}
The coefficients obtained from the fitting procedure are used to reconstruct the correlation function onto the points along the parametrized curves with no measured data, as shown in panels (b) and (e) of Fig.~\ref{fig:fit}.

\begin{figure}[ht]
    \centering
    \includegraphics[width=0.95\linewidth]{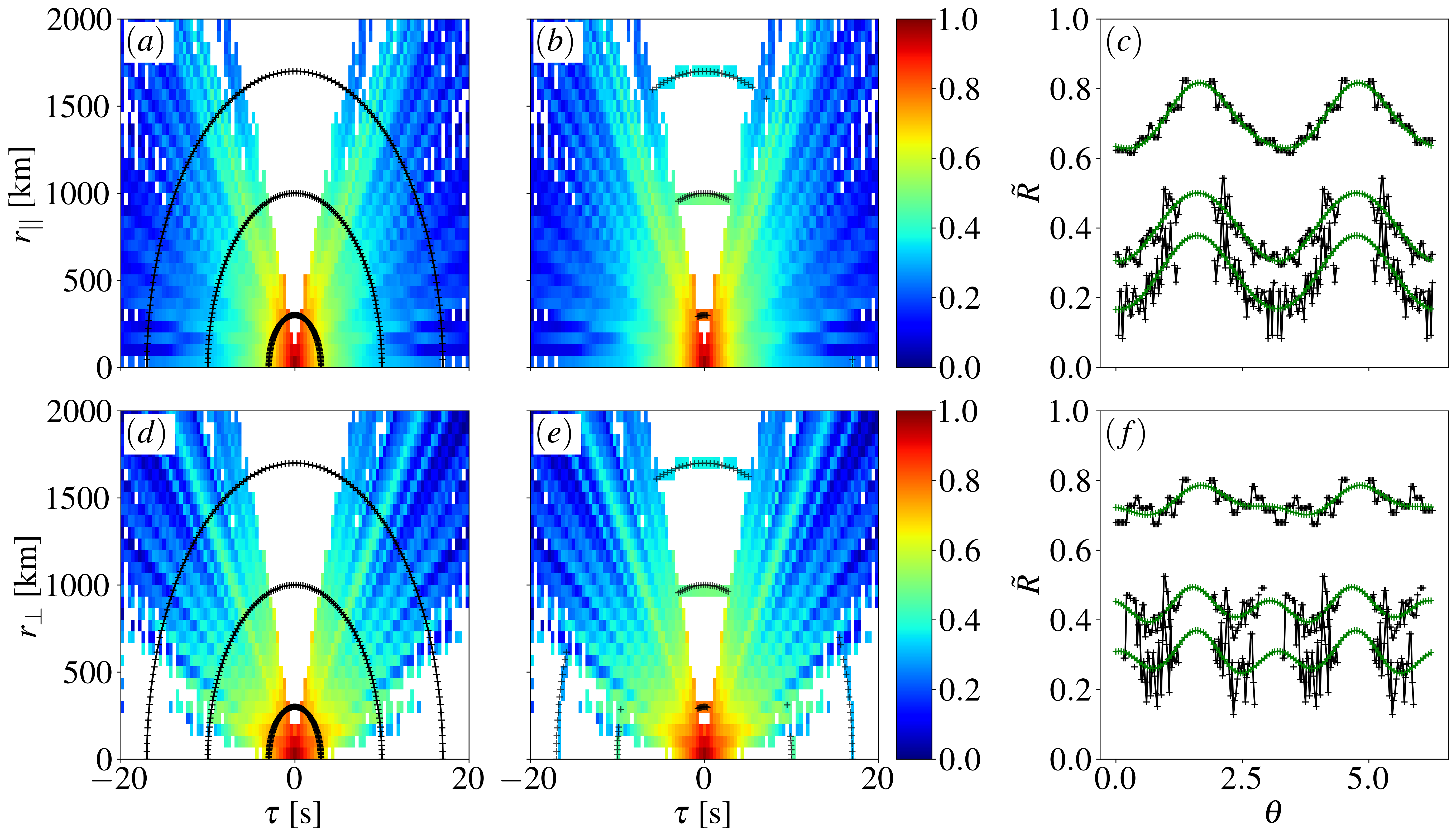}
    \caption{Fitting procedure to obtain complete spatiotemporal coverage. Top row shows the parallel correlation function, bottom row the perpendicular. A few parametric curves (3 black curves) are indicated in (a) and (d) on top of the measured correlation functions. The filled-in points are evidenced in the middle panels, (b) and (e), as colored values with a ``+'' sign on top. The fitting result used to reproduce the correlation function profiles is shown in (c) and (f).}
    \label{fig:fit}
\end{figure}

The full procedure involves 35 parametric curves, spaced by steps of 50~km. A few of these curves and the fitted continuous function in Eq.~\ref{eq:fit} are shown in Fig.~\ref{fig:fit} (c) and (f). As a result of this procedure, most of the unmapped regions of the plane are covered, as shown in Fig.~\ref{fig:R_fit}

\begin{figure}[ht]
    \centering
    \includegraphics[width=0.95\linewidth]{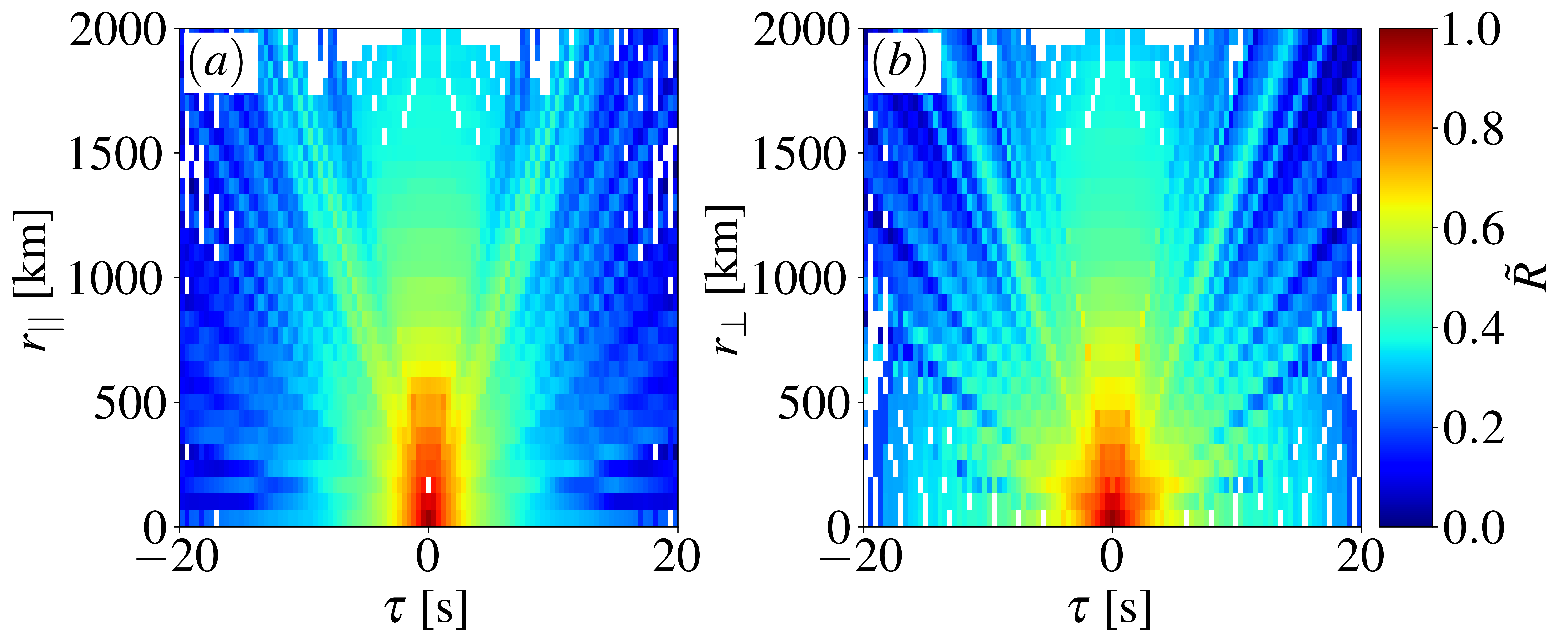}
    \caption{Space-time correlation function with enhanced coverage obtained from the fitting procedure.}
    \label{fig:R_fit}
\end{figure}

Finally, the few missing white spots that were not covered by any parametric curve are filled in using a linear interpolation algorithm to obtain the correlation function in Fig.~\ref{fig:R_fit_interp} over the full space-time domain.
}

\showmatmethods{} % Display the Materials and Methods section

\acknow{We dedicate this paper
to the memory of 
co-author Prof. Vincenzo Carbone, long-time colleague and 
key contributor to the 
conceptualization of this project, who passed away unexpectedly in January 2025.
This work is supported by NSF grant PHY-2108834, and by NASA MMS Mission under grant number 80NSSC19K0565 at the University of Delaware, by subcontract SUB0000517 from Princeton funded by NASA MMS HGIO project 80NSSC21K0739, by a plan for NASA EPSCoR Research Infrastructure Development (RID) in Delaware (NASA award 80NSSC22M0039), and by the Space It Up project funded by the Italian Space Agency, ASI, and the Ministry of University and Research, MUR, under contract n. 2024-5-E.0 - CUP n. I53D24000060005. This study was conducted using the numerical facility at the Delaware Space Observation Center (DSpOC), which is supported by NASA Awards 80NSSC22K0884 and 80NSSC23K0608.}

\showacknow{} % Display the acknowledgments section

%\bibsplit[2] %#@
%Use \bibsplit to split the references from the body of the text. Value "[2]" represents the number of reference in the left column (Note: Please avoid single column figures & tables on this page.)

% Bibliography
%\bibliography{biblio}

\begin{thebibliography}{10}

\bibitem{chen1989sweeping}
S Chen, RH Kraichnan, Sweeping decorrelation in isotropic turbulence.
\newblock {\em\protect\JournalTitle{Physics of Fluids A: Fluid Dynamics}} \textbf{1}, 2019--2024 (1989).

\bibitem{tennekes1975eulerian}
H Tennekes, Eulerian and lagrangian time microscales in isotropic turbulence.
\newblock {\em\protect\JournalTitle{Journal of Fluid Mechanics}} \textbf{67}, 561–567 (1975).

\bibitem{taylor1938spectrum_frozenin}
GI Taylor, The spectrum of turbulence.
\newblock {\em\protect\JournalTitle{Proceedings of the Royal Society of London. Series A, Mathematical and Physical Sciences}} \textbf{164}, 476--490 (1938).

\bibitem{fox2016solar}
NJ Fox, et~al., The solar probe plus mission: humanity’s first visit to our star.
\newblock {\em\protect\JournalTitle{Space Science Reviews}} \textbf{204}, 7--48 (2016).

\bibitem{klein2015modified}
KG Klein, JC Perez, D Verscharen, A Mallet, BDG Chandran, A modified version of taylor’s hypothesis for solar probe plus observations.
\newblock {\em\protect\JournalTitle{The Astrophysical Journal Letters}} \textbf{801}, L18 (2015).

\bibitem{bourouaine2018limitations}
S Bourouaine, JC Perez, On the limitations of taylor’s hypothesis in parker solar probe’s measurements near the alfvén critical point.
\newblock {\em\protect\JournalTitle{The Astrophysical Journal Letters}} \textbf{858}, L20 (2018).

\bibitem{chhiber2019contextual}
R Chhiber, AV Usmanov, WH Matthaeus, TN Parashar, ML Goldstein, Contextual predictions for parker solar probe. {II}. turbulence properties and taylor hypothesis.
\newblock {\em\protect\JournalTitle{The Astrophysical Journal Supplement Series}} \textbf{242}, 12 (2019).

\bibitem{perez2021applicability}
JC Perez, S Bourouaine, CHK Chen, NE Raouafi, Applicability of taylor’s hypothesis during parker solar probe perihelia.
\newblock {\em\protect\JournalTitle{A\&A}} \textbf{650}, A22 (2021).

\bibitem{matthaeus2010eulerian}
WH Matthaeus, S Dasso, JM Weygand, MG Kivelson, KT Osman, Eulerian decorrelation of fluctuations in the interplanetary magnetic field.
\newblock {\em\protect\JournalTitle{The Astrophysical Journal Letters}} \textbf{721}, L10 (2010).

\bibitem{burch2016MMS}
JL Burch, TE Moore, RB Torbert, BL Giles, Magnetospheric multiscale overview and science objectives.
\newblock {\em\protect\JournalTitle{Space Science Reviews}} \textbf{199}, 5--21 (2016).

\bibitem{klein2023helioswarm}
KG Klein, et~al., Helioswarm: A multipoint, multiscale mission to characterize turbulence.
\newblock {\em\protect\JournalTitle{Space Science Reviews}} \textbf{219}, 74 (2023).

\bibitem{retino2022particle_PO}
A Retin{\`o}, et~al., Particle energization in space plasmas: towards a multi-point, multi-scale plasma observatory.
\newblock {\em\protect\JournalTitle{Experimental Astronomy}} \textbf{54}, 427--471 (2022).

\bibitem{marcucci2024PO}
MF {Marcucci}, A {Retin{\`o}}, {The ESA M7 candidate mission Plasma Observatory.} in {\em EGU General Assembly Conference Abstracts}, EGU General Assembly Conference Abstracts.
\newblock p. 11903 (2024).

\bibitem{batchelor1953theory}
GK Batchelor, {\em The theory of homogeneous turbulence}.
\newblock (Cambridge university press), (1953).

\bibitem{MoninYaglomI}
AS Monin, AM Yaglom, {\em Statistical Fluid Mechanics, Vol. 1}.
\newblock (MIT Press, Cambridge, Mass.), (1971).

\bibitem{MoninYaglomII}
AS Monin, AM Yaglom, {\em Statistical Fluid Mechanics, Vol. 2}.
\newblock (MIT Press, Cambridge, Mass.), (1975).

\bibitem{zhou2004colloquium}
Y Zhou, WH Matthaeus, P Dmitruk, Colloquium: Magnetohydrodynamic turbulence and time scales in astrophysical and space plasmas.
\newblock {\em\protect\JournalTitle{Rev. Mod. Phys.}} \textbf{76}, 1015--1035 (2004).

\bibitem{bieber1994proton}
JW {Bieber}, et~al., {Proton and Electron Mean Free Paths: The Palmer Consensus Revisited}.
\newblock {\em\protect\JournalTitle{The Astrophysical Journal}} \textbf{420}, 294 (1994).

\bibitem{shalchi2006parallel}
A Shalchi, JW Bieber, WH Matthaeus, R Schlickeiser, Parallel and perpendicular transport of heliospheric cosmic rays in an improved dynamical turbulence model.
\newblock {\em\protect\JournalTitle{The Astrophysical Journal}} \textbf{642}, 230 (2006).

\bibitem{ridley2000estimations}
A Ridley, Estimations of the uncertainty in timing the relationship between magnetospheric and solar wind processes.
\newblock {\em\protect\JournalTitle{Journal of Atmospheric and Solar-Terrestrial Physics}} \textbf{62}, 757--771 (2000).

\bibitem{orszag1977lectures}
SA Orszag, Lectures on the statistical theory of turbulence, eds.{} R Balian, JL Peube.
\newblock (Gordon and Breach, New York), p. 235 (1977) Les Houches Summer School, 1973.

\bibitem{roy2022turbulent}
S Roy, et~al., Turbulent energy transfer and proton–electron heating in collisionless plasmas.
\newblock {\em\protect\JournalTitle{The Astrophysical Journal}} \textbf{941}, 137 (2022).

\bibitem{edwards1964statistical}
SF Edwards, The statistical dynamics of homogeneous turbulence.
\newblock {\em\protect\JournalTitle{Journal of Fluid Mechanics}} \textbf{18}, 239–273 (1964).

\bibitem{orszag1970analytical}
SA Orszag, Analytical theories of turbulence.
\newblock {\em\protect\JournalTitle{Journal of Fluid Mechanics}} \textbf{41}, 363–386 (1970).

\bibitem{orszag1972numerical}
SA {Orszag}, GS {Patterson}, {Numerical Simulation of Three-Dimensional Homogeneous Isotropic Turbulence}.
\newblock {\em\protect\JournalTitle{Physical Review Letters}} \textbf{28}, 76--79 (1972).

\bibitem{zhou2021turbulence}
Y Zhou, Turbulence theories and statistical closure approaches.
\newblock {\em\protect\JournalTitle{Physics Reports}} \textbf{935}, 1--117 (2021).

\bibitem{comtebellot1971simple}
G Comte-Bellot, S Corrsin, Simple eulerian time correlation of full-and narrow-band velocity signals in grid-generated, ‘isotropic’ turbulence.
\newblock {\em\protect\JournalTitle{Journal of Fluid Mechanics}} \textbf{48}, 273–337 (1971).

\bibitem{servidio2011time}
S {Servidio}, V {Carbone}, P {Dmitruk}, WH {Matthaeus}, {Time decorrelation in isotropic magnetohydrodynamic turbulence}.
\newblock {\em\protect\JournalTitle{EPL (Europhysics Letters)}} \textbf{96}, 55003 (2011).

\bibitem{lugones2016spatiotemporal}
R Lugones, P Dmitruk, PD Mininni, M Wan, WH Matthaeus, On the spatio-temporal behavior of magnetohydrodynamic turbulence in a magnetized plasma.
\newblock {\em\protect\JournalTitle{Physics of Plasmas}} \textbf{23}, 112304 (2016).

\bibitem{lugones2019spatiotemporal}
R Lugones, P Dmitruk, PD Mininni, A Pouquet, WH Matthaeus, Spatio-temporal behavior of magnetohydrodynamic fluctuations with cross-helicity and background magnetic field.
\newblock {\em\protect\JournalTitle{Physics of Plasmas}} \textbf{26}, 122301 (2019).

\bibitem{kolmogorov1941local}
AN Kolmogorov, {The Local Structure of Turbulence in Incompressible Viscous Fluid for Very Large Reynolds' Numbers}.
\newblock {\em\protect\JournalTitle{Proc. R. Soc. Lond.}} \textbf{434}, 9–13 (1941).

\bibitem{iroshnikov1964turbulence}
PS {Iroshnikov}, {Turbulence of a Conducting Fluid in a Strong Magnetic Field}.
\newblock {\em\protect\JournalTitle{Soviet Ast.}} \textbf{7}, 566 (1964).

\bibitem{kraichnan1965inertial}
RH Kraichnan, Inertial‐range spectrum of hydromagnetic turbulence.
\newblock {\em\protect\JournalTitle{The Physics of Fluids}} \textbf{8}, 1385--1387 (1965).

\bibitem{kraichnan1959structure}
RH Kraichnan, The structure of isotropic turbulence at very high reynolds numbers.
\newblock {\em\protect\JournalTitle{Journal of Fluid Mechanics}} \textbf{5}, 497–543 (1959).

\bibitem{kraichnan1965lhdia}
RH Kraichnan, Lagrangian‐history closure approximation for turbulence.
\newblock {\em\protect\JournalTitle{The Physics of Fluids}} \textbf{8}, 575--598 (1965).

\bibitem{sanada1992random}
T {Sanada}, V {Shanmugasundaram}, {Random sweeping effect in isotropic numerical turbulence}.
\newblock {\em\protect\JournalTitle{Physics of Fluids A}} \textbf{4}, 1245--1250 (1992).

\bibitem{mccomb1989velocity}
WD Mccomb, V Shanmugasundaram, P Hutchinson, Velocity-derivative skewness and two-time velocity correlations of isotropic turbulence as predicted by the let theory.
\newblock {\em\protect\JournalTitle{Journal of Fluid Mechanics}} \textbf{208}, 91–114 (1989).

\bibitem{zhou2024hydrodynamic}
Y Zhou, {\em Hydrodynamic Instabilities and Turbulence: Rayleigh–Taylor, Richtmyer–Meshkov, and Kelvin–Helmholtz Mixing}.
\newblock (Cambridge University Press), (2024).

\bibitem{brunocarbone2005review}
R Bruno, V Carbone, The solar wind as a turbulence laboratory.
\newblock {\em\protect\JournalTitle{Living Reviews in Solar Physics}} \textbf{2}, 4 (2005).

\bibitem{frisch1995turbulence}
U {Frisch}, {\em {Turbulence: the legacy of AN Kolmogorov}}.
\newblock (Cambridge university press), (1995).

\bibitem{lumley1965interpretation}
JL Lumley, Interpretation of time spectra measured in high‐intensity shear flows.
\newblock {\em\protect\JournalTitle{The Physics of Fluids}} \textbf{8}, 1056--1062 (1965).

\bibitem{jokipii1973turbulence}
JR Jokipii, Turbulence and scintillations in the interplanetary plasma.
\newblock {\em\protect\JournalTitle{Annual Review of Astronomy and Astrophysics}} \textbf{11}, 1--28 (1973).

\bibitem{matthaeus2016ensemble}
WH Matthaeus, JM Weygand, S Dasso, Ensemble space-time correlation of plasma turbulence in the solar wind.
\newblock {\em\protect\JournalTitle{Phys. Rev. Lett.}} \textbf{116}, 245101 (2016).

\bibitem{pecora2025ubmsh}
F Pecora, et~al., General properties of the magnetosheath: the mms unbiased campaign.
\newblock {\em\protect\JournalTitle{Submitted to JGR: Space Physics}} (2025).

\bibitem{wang2024anisotropy}
J Wang, et~al., Anisotropy of density fluctuations in the solar wind at 1 au.
\newblock {\em\protect\JournalTitle{The Astrophysical Journal}} \textbf{967}, 150 (2024).

\bibitem{pecora2023threedimensional_PRL}
F Pecora, et~al., Three-dimensional energy transfer in space plasma turbulence from multipoint measurement.
\newblock {\em\protect\JournalTitle{Phys. Rev. Lett.}} \textbf{131}, 225201 (2023).

\bibitem{huang2017existence}
SY Huang, LZ Hadid, F Sahraoui, ZG Yuan, XH Deng, On the existence of the kolmogorov inertial range in the terrestrial magnetosheath turbulence.
\newblock {\em\protect\JournalTitle{The Astrophysical Journal Letters}} \textbf{836}, L10 (2017).

\bibitem{sahraoui2020magnetohydrodynamic}
F Sahraoui, L Hadid, S Huang, Magnetohydrodynamic and kinetic scale turbulence in the near-earth space plasmas: a (short) biased review.
\newblock {\em\protect\JournalTitle{Reviews of Modern Plasma Physics}} \textbf{4}, 4 (2020).

\bibitem{azelis2024spacetime}
AA Azelis, JC Perez, S Bourouaine, Space–time structure of weak magnetohydrodynamic turbulence.
\newblock {\em\protect\JournalTitle{Journal of Plasma Physics}} \textbf{90}, 905900109 (2024).

\bibitem{shebalin1983anisotropy}
JV Shebalin, WH Matthaeus, D Montgomery, Anisotropy in mhd turbulence due to a mean magnetic field.
\newblock {\em\protect\JournalTitle{Journal of Plasma Physics}} \textbf{29}, 525–547 (1983).

\bibitem{oughton1994influence}
S Oughton, ER Priest, WH Matthaeus, The influence of a mean magnetic field on three-dimensional magnetohydrodynamic turbulence.
\newblock {\em\protect\JournalTitle{Journal of Fluid Mechanics}} \textbf{280}, 95–117 (1994).

\bibitem{oughton2006twocomponent}
S Oughton, P Dmitruk, WH Matthaeus, A two-component phenomenology for homogeneous magnetohydrodynamic turbulence.
\newblock {\em\protect\JournalTitle{Physics of Plasmas}} \textbf{13}, 042306 (2006).

\bibitem{shue1997new}
JH Shue, et~al., A new functional form to study the solar wind control of the magnetopause size and shape.
\newblock {\em\protect\JournalTitle{Journal of Geophysical Research: Space Physics}} \textbf{102}, 9497--9511 (1997).

\bibitem{russell2016magnetospheric}
CT Russell, et~al., The magnetospheric multiscale magnetometers.
\newblock {\em\protect\JournalTitle{Space Science Reviews}} \textbf{199}, 189--256 (2016).

\bibitem{pecora2023relaxation}
F Pecora, et~al., {Relaxation of the turbulent magnetosheath}.
\newblock {\em\protect\JournalTitle{Monthly Notices of the Royal Astronomical Society}} \textbf{525}, 67--72 (2023).

\end{thebibliography}

\end{document}